\documentstyle[eqsecnum,preprint,aps,prd,epsfig]{revtex}
\preprint{WISC-MILW-94-TH-19} 
\begin{document}
\tighten
\draft

\title{ LONG RANGE EFFECTS \\ OF COSMIC STRING STRUCTURE }

\author{Bruce Allen} \address{ Department of Physics, University of
Wisconsin -- Milwaukee\\ P.O. Box 413, Milwaukee, Wisconsin 53201,
U.S.A.\\ email: ballen@dirac.phys.uwm.edu}

\author{Bernard S. Kay} \address{Department of Mathematics, University
of York,\\ Heslington, York YO1 5DD, U.K\\ email:
bsk2@unix.york.ac.uk}

\author{Adrian C. Ottewill} \address{Department of Mathematical
Physics, University College Dublin,\\ Belfield, Dublin 4, Ireland\\
email: ottewill@relativity.ucd.ie}

\maketitle
\begin{abstract}
   We combine and further develop ideas and techniques of Allen \&
Ottewill, Phys.~Rev.~D, {\bf 42}, 2669 (1990) and Kay \& Studer
Commun.~Math.~Phys., {\bf 139}, 103 (1991) for calculating the long
range effects of cosmic string cores on classical and quantum field
quantities far from an (infinitely long, straight) cosmic string.  We
find analytical approximations for (a) the gravity-induced ground
state renormalized expectation values of $\hat\varphi^2$ and $\hat
T_\mu{}^\nu$ for a non-minimally coupled quantum scalar field far from
a cosmic string (b) the classical electrostatic self
force on a test charge far from a superconducting cosmic string.
Surprisingly -- even at cosmologically large distances -- all these
quantities would be very badly approximated by idealizing the string
as having zero thickness and imposing regular boundary conditions; 
instead they are well approximated by suitably fitted strengths
of logarithmic divergence at the string core. 
Our formula for ${\langle {\hat \varphi}^2 \rangle}$
reproduces (with much less effort and much more generality) the
earlier numerical results of Allen \& Ottewill.  Both ${\langle 
{\hat \varphi}^2 \rangle}$ and
${\langle {\hat T}_{\mu}{}^{\nu} \rangle}$
 turn out to be ``weak field topological invariants''
depending on the details of the string core only through the minimal
coupling parameter ``$\xi$'' (and the deficit angle).  Our formula for
the self-force (leaving aside relatively tiny gravitational
corrections) turns out to be attractive: We obtain, for the
self-potential of a test charge $Q$ a distance $r$ from a (GUT scale)
superconducting string, the formula $- Q^2/(16\epsilon_0r\ln(qr))$
where $q$ is an (in principle, computable) constant of the order of 
the inverse string radius.
\end{abstract}

\section{INTRODUCTION}
\label{intro}

A realistic cosmic string has structure on a length scale defined by
the phase transition at which it is formed. In the case of a GUT
string this corresponds to a radius of order $10^{-30}\,$cm.  As this
radius is so small, one often models the true string space-time by an
idealized space-time where the string core has zero thickness and the
curvature is represented by a 2-dimensional delta-function.  The
idealized model for an infinitely long straight static cosmic string
space-time is the manifold $ R^2 \times  R^+ \times  S^1$ 
with conical metric
\begin{equation}
  {\rm d}s^2 = - {\rm d}t^2 + {\rm d}z^2 +{\rm d}r^2 +r^2{\rm d}\phi^2
                                      \quad ,\label{loridealds2}
\end{equation}
where the angular range is $\phi \in [0,2\pi/\kappa)$, corresponding
to a deficit angle of $2\pi (1 - 1/\kappa)$.  Throughout this paper we
shall assume the standard case of positive deficit angle, so that
$\kappa >1$; for GUT strings $\kappa -1 \sim 10^{-6}$.

  In studying the behaviour of various types of fields and waves far
from such a string, it is sometimes erroneously taken for granted that
one can always ignore the details of the interior structure of the
string and approximate the effect of the string by this idealized
model with regularity conditions placed at the conical singularity.
This paper presents two calculations involving fields propagating
around and interacting with a cosmic string for which this is not
true.  These calculations have, instead, the remarkable feature that
the quantity calculated, even at very large distances from the string,
depends on details of the interaction inside the string core.

 We shall start by treating a `realistic' or `true' cosmic string
having a core of finite thickness with a definite radius $a$ (but
assumed for simplicity to be infinitely long, straight and static).
This corresponds to a space-time metric taking the standard conical
form with given deficit angle outside the radius $a$, but matching
onto a smooth model core metric inside this radius (see
Eq. (\ref{ds2}) below).

The first calculation involves a non-minimally coupled {\it quantum}
linear scalar field $\hat\varphi(x)$: We shall obtain an approximate
formula for the renormalized vacuum expectation value ${\langle {\hat \varphi}^2 \rangle}$ for
such a field far from the string and a similar formula for the
expectation value of the energy-momentum tensor 
${\langle {\hat T}_{\mu}{}^{\nu} \rangle}$.  The
second calculation concerns a {\it classical} electrostatic field at
large distances from a {\it superconducting} cosmic string: In this
case the string is additionally characterized by a non-vanishing
function within the core representing the `local photon mass term'
responsible for making it superconducting \cite{witten}.  We shall
obtain an approximate formula for the self-force on a test charge far
from the string due to the presence of such a string.  (As we shall
see, this self-force arises as the sum of two terms: A small repulsive
term previously calculated by Smith \cite{smith} and Linet
\cite{linet} which depends on the deficit angle and a typically much
larger attractive term which depends on the ``scattering length'' of
the local photon mass term and is independent of the deficit angle.)
 
Both calculations presented here involve calculating Green functions
$G$ for equations of the schematic form
\begin{equation}
(-\Delta + V)G=\delta \label{generic}
\end{equation}
and both involve calculating (suitably renormalized) coincident-point
values of such Green functions. In the calculation of 
${\langle {\hat \varphi}^2 \rangle}$,
$\Delta$ represents the Laplace-Beltrami operator for the
(Euclideanized) 4-dimensional string space-time metric, while $V$
represents a non-minimal coupling term $\xi {\cal R}$, where $\cal R$
is the Ricci scalar of the same metric.  In the calculation of the
self-force, $\Delta$ represents the Laplace-Beltrami operator for the
3-dimensional spatial metric of the string at a fixed time while $V$
represents the local photon mass term regarded as a function on that
3-dimensional space.  For both calculations presented here, the
sensitivity to the core structure can be traced back to the potential
term $V$ in Eq.\ (\ref{generic}).  For example, in the case of the
calculation of ${\langle {\hat \varphi}^2 \rangle}$ for a {\it
  minimally} coupled quantum
scalar field (corresponding to $V =0$), one can idealize the true
space-time of the cosmic string by the idealized conical space-time
(\ref{loridealds2}) and, with the imposition of regularity conditions
on the scalar field at the conical singularity, obtain an excellent
approximation to ${\langle {\hat \varphi}^2 \rangle}$ at large
distances from the string (seeSec. \ref{expectation}).

That the value of ${\langle {\hat \varphi}^2 \rangle}$ for a 
{\it non-minimally} coupled field
will depend on the details of the metric in the core of the string
even very far from the string was argued by Allen and Ottewill in
\cite{AO}. There the argument was confirmed by detailed calculations
for two model cores: the `flower pot' and the `ball-point pen'.  In
particular, the value of ${\langle {\hat \varphi}^2 \rangle}$ for 
the flower pot model was
computed numerically and shown to differ significantly from the ideal
value out to cosmological scales for GUT scale strings.

Roughly simultaneously with the work of Allen and Ottewill, Kay and
Studer \cite{KS} looked at the question of boundary conditions at the
conical singularity for a variety of situations involving (classical
and quantum) scalar fields and waves around an idealized string.  They
found that there is typically a one-parameter family of possible
boundary conditions for the idealized problem -- one of which is
regular and the others of which involve a field which, at each time
$t$ is logarithmically divergent near the origin:

\begin{equation}
\varphi \sim {\rm const}{\cdot}\ln(r/R) , \label{kayasymp}
\end{equation}
where $R$ is a quantity with the dimensions of length labelling the
boundary condition \cite{footnote}.
Moreover, they argued that, in the case of many physical quantities
involving such a field around a true string, and in particular in the
case of Eq.\ (\ref{generic}), one should be able to well approximate
the effect of the string core by a single parameter with the
dimensions of length which they introduced and termed the
`2-dimensional scattering length' \cite{KS}.  This length is easily
determined in terms of the core metric and $V$ by what they termed
their `fitting formula' (Eq.\ (5.9) of \cite{KS} and equation
(\ref{fitting}) here).  The approximation proceeds by idealizing the
string, but rather than taking regularity conditions at the conical
singularity, imposing the boundary condition (\ref{kayasymp}) where
$R$ is identified with this scattering length.  Only in cases where
$V$ in Eq. (\ref{generic}) vanishes, when one can show that the
scattering length $R$ will be automatically zero, will it be justified
to approximate the true string by the idealized string with regular
boundary conditions.  Non-vanishing $V$ will in general give rise to
non-vanishing scattering lengths and hence require approximation by
idealized strings with non-regular (i.e.  suitably logarithmically
diverging) boundary conditions.  In the
present paper, we shall always assume $V$ to be non-negative, 
and, in consequence,
it may easily be shown that the corresponding scattering lengths $R$,
while non-vanishing, will necessarily be small (bounded by the string
radius $a$ in all cases, and, for the problem of
 ${\langle {\hat \varphi}^2 \rangle}$, even
``exponentially small'').  Nevertheless, and quite surprisingly, we
shall find that the failure of $R$ to be precisely zero makes a big
difference to the effects we calculate, even at cosmologically large
scales and it will turn out to be crucially important, if one
idealizes the string as having zero thickness, to take the appropriate
logarithmically divergent boundary conditions, rather than regular
boundary conditions in order to obtain valid approximations.

Kay and Studer speculated that this method of fitting the idealized
boundary condition to the true scattering length may lead to a useful
analytical approximation to Allen and Ottewill's calculations of
${\langle {\hat \varphi}^2 \rangle}$ for a non-minimally coupled
scalar field (see the end of
Sec.\ 5 of \cite{KS}).  They also discussed how this procedure could
be used to approximate the scattering theory of electromagnetic fields
by superconducting cosmic strings. Furthermore, they also pointed out
(see Note 16 of \cite{KS}) that the self-potential of a test charge
due to the presence of a cosmic string (previously calculated by Smith
\cite{smith} and Linet \cite{linet} who only took into account the
effect of a conical geometry) should have an additional important
contribution from the local photon mass term in the case the string
were superconducting. They again suggested that this might be well
approximated analytically by replacing the true problem (i.e. with the
photon mass term) by the problem of finding the electrostatic
potential (i.e. Green function) due to a point test charge in the
presence of an idealized cosmic string when the potential is obliged
to satisfy the boundary condition (\ref{kayasymp}) at the string and
then taking the appropriate renormalized coincidence limit.

However, it turns out that when one tries to pursue these ideas to
obtain approximate analytical formulae for 
${\langle {\hat \varphi}^2 \rangle}$ and the
self-force one encounters certain difficulties, as was partly
anticipated in \cite{KS}. These difficulties are associated with the
fact that the idealized problem with non-zero $R$ will have a bound
state, which however for small scattering lengths (less than or around
$a$ as will be the case here) is not expected to be `believable' (see
\cite{KS} and \cite{KF}). If one attempts to implement the proposals
in \cite{KS} literally, this is reflected in the existence of spurious
poles in certain integrals (see, for example, Eq.~(\ref{correction})
in the present paper).

In the present paper we show, by a combination of ideas and techniques
derived both from \cite{AO} and \cite{KS}, that such suitable
modifications can be made. We then obtain approximate analytical
formulae both for ${\langle {\hat \varphi}^2 \rangle}$ 
(Eq.~(\ref{approxphi2})) and for the
self-force (Eqs.~(\ref{Fsuper})) which depend on $V$ only through its
fitted scattering length $R$, and which, in the case of
 ${\langle {\hat \varphi}^2 \rangle}$
give an excellent approximation to the numerical results of \cite{AO}.
In this way, the basic philosophy of \cite{KS} is vindicated.

Remarkably, for small deficit angle we find that the scattering length
$R$ required to approximate the calculation of 
${\langle {\hat \varphi}^2 \rangle}$ is a ``weak
field topological invariant'' given by
\[
R \approx a \exp \left( - {1 \over 2 \xi (\kappa - 1) } \right) .
\]
Thus, in this case ${\langle {\hat \varphi}^2 \rangle}$ is actually 
insensitive to the detailed
shape of the string core and depends on the interaction with the
string core only through the single ``non-minimal coupling'' parameter
$\xi$ and the deficit angle.

We remark that, as discussed in \cite{KS}, in the self-force problem,
the typical values
for the scattering length $R$ of the local photon mass term
are expected to be of the order of the string radius $a$.  While, at
first sight, this is ``very small'' compared to the distances of
interest, as we have already anticipated above (and as, in this case,
was already anticipated in \cite{KS} -- see ``Pitfall 2'' in Note 22
there) one can argue that such values of $R$ will lead to effects at
``medium scales'' which are significantly different from the effects
one would calculate in the case $R$ were precisely zero. (The reason
for this is essentially because $R$ and the scales of interest are
expected to occur -- because of the ``2-dimensional'' nature of the
problem -- in the combination $\ln({\rm scale}/R)$.)  This is borne
out in the present paper by our self-force calculation.
More spectacularly than this, for the problem of 
${\langle {\hat \varphi}^2 \rangle}$, we shall
see that typical $R$ values will be {\it exponentially} small compared
to $a$, ($R \approx {\rm e}^{-3,000,000} a$ for a GUT scale string
with $\xi=1/6$).  Yet, we shall continue to find (and this goes beyond
anything envisaged in \cite{KS}) that the corresponding value of
${\langle {\hat \varphi}^2 \rangle}$ differs significantly from 
the value one would obtain in
the case $R$ were precisely zero (i.e. from what one would obtain in
the case of naive regularity conditions) even on cosmological scales.
Thus we conclude, as we have already mentioned, that, for this problem
too, even though the scattering lengths $R$ are so incredibly tiny, it
continues to be important not to replace them by zero!

Throughout this paper we shall work with a positive definite
metric. This is a valid and convenient way to treat the quantum field
theory since the space-time is static \cite{wald} (and is irrelevant
to the classical self-force calculation).  It is however a crucial step in
our approach since it leads to computations of Green functions that
fall off rapidly in all directions from the string.  By replacing the
Lorentzian signature metric with a positive definite one, the
hyperbolic problem for $G$ becomes an elliptic one, with a unique
regular solution $G$ which falls off in all directions away from the string.

\section{GREEN FUNCTIONS}
\label{greenfunctions}

In Allen \& Ottewill\cite{AO}, the quantum field theory of a scalar
field was studied on a model string space-time in which the string was
still taken to be infinitely long, straight and static but the core of
the string was given a non-zero spatial extent characterised by a
length scale $a$.  The (positive definite) metric was written in the
form
\begin{equation}
 {\rm d}s^2= {\rm d}t^2+{\rm d}z^2+P^2(r/a) {\rm d}r^2+r^2 {\rm
                                              d}\phi^2, \label{ds2}
\end{equation}
where the range of the angular coordinate is $\phi \in
[0,2\pi/\kappa)$, and $P(r/a)$ is a smooth monotonic function
satisfing the equations
\begin{equation}
 \lim_{r /a \to 0} P(r/a) = 1/\kappa \quad {\rm and} \quad P(r/a) = 1
 \qquad r>a.
\end{equation}
The first condition means that there is no conical singularity at
$r=0$. The second condition means that the curvature is confined
within a cylinder of radius $a$, the string core, and that, viewed
from outside this core, the space has the standard deficit angle $2\pi
(1 - 1/\kappa)$.  The second condition here is actually slightly
stronger than that used in Ref. \cite{AO} but agrees with the
condition used by Kay \& Studer\cite{KS} and is more convenient for
our purposes here.  (Note the unfortunate clash of notation that
$\kappa$ as defined in \cite{KS} is the inverse of the $\kappa$ as
defined in \cite{AO}.  We follow the latter convention here so that a
positive deficit angle corresponds to $\kappa \in (1,\infty)$.)

We wish to construct Green functions for the scalar `wave equation' on
this space-time and Laplace's equation on its constant $t$ sections
with positive cylindrically symmetric potential $V$ whose support lies
in $r \leq a$.  These Green functions satisfy
\begin{equation}
\left(- \Box + {1 \over a^2}V(r/a)\right)G^{(4)}(x,x') =
      \delta^{(4)}(x,x'), \label{potentialeqn}
\end{equation}
where $\Box$ is the Laplace-Beltrami operator for the metric
(\ref{ds2}) and $\delta^{(4)}$ is the 4-dimensional covariant
delta-function, and
\begin{equation}
\left(- \triangle + {1 \over a^2}V(r/a)\right)G^{(3)}({\bf x},{\bf
                        x'}) = \delta^{(3)}({\bf x},{\bf x'}),
                        \label{3potentialeqn}
\end{equation}
where $\triangle$ is the Laplace-Beltrami operator for a constant $t$
section and $\delta^{(3)}$ is the 3-dimensional covariant
delta-function.  We have written the potential in this form so that
(a) $V$ is a dimensionless function of a dimensionless argument
$x=r/a$ and (b) its integral over a spatial slice of constant $z$ is
independent of $a$.

Eq.\ (\ref{potentialeqn}) includes the case of a scalar field with
curvature coupling $\xi \ge 0$, if one identifies $V(x)=\xi {\cal
R}(x)$, since we then have

\begin{equation}
       \xi{\cal R}(x)={\xi \over a^2} { 2 \over (r/a)} {P'(r/a) \over
P^3(r/a)} \quad = {2 \xi \over x} {P'(x) \over P^3(x)}
\label{ricci}
\end{equation}
 which is positive as we have assumed $P(r/a)$ to be a monotone
increasing function.  Here and throughout, the notation $f'$ denotes
the derivative of the function $f$ with respect to its argument.
Equation (\ref{3potentialeqn}) is of interest to us as the equation
for the electrostatic potential on a superconducting string. Here, 
stability requires \cite{witten} the absence of ``bound states'' for the 
Schr\"odinger-like operator in (\ref{3potentialeqn}) and, for
simplicity, we shall take the potential $V$ to be everywhere non-negative.

The homogeneous form of Eq. (\ref{potentialeqn}) admits solutions of
the form
\begin{equation}
 e^{i \omega t} e^{i k z} e^{in\kappa \phi} \Psi_n(r/a;sa),
                   \label{solns}
\end{equation}
where $\omega$, $k \in  R$, $s^2 \equiv \omega^2 + k^2$, $n \in
{ Z}$ and $\Psi_n(r/a;sa)$ satisfies
\begin{equation}
 \biggl[ - {1 \over xP(x)}{{\rm d} \over {\rm d}x} {x \over P(x)}
{{\rm d} \over {\rm d}x} + (sa)^2 + {n^2 \kappa^2 \over x^2} + V(x)
\biggr] \Psi_n(x;sa) = 0 .  \label{radialeqn}
\end{equation}
The Green function for the `wave equation' may then be written as
\begin{eqnarray}
 G^{(4)}(x,x') &=& \int\limits_{-\infty}^\infty \> {{\rm d}\omega
                     \over 2\pi} e^{i \omega \Delta t}
                     \int\limits_{-\infty}^\infty \> {{\rm d}k \over
                     2\pi} e^{i k \Delta z}
                     \sum_{n=-\infty}^{\infty}{\kappa \over 2 \pi } \>
                     e^{in\kappa \Delta \phi} \Psi_n^{\,<}(r_</a;sa)
                     \Psi_n^{\,>}(r_>/a;sa) , \\
     &=& {\kappa \over 4 \pi^2} \int\limits_0^\infty \> s {\rm d}s \>
                     J_0\left(s \sqrt{\Delta t^2 + \Delta z^2}\right)
                     \sum_{n=-\infty}^{\infty} e^{in\kappa \Delta
                     \phi} \Psi_n^{\,<}(r_</a;sa)
                     \Psi_n^{\,>}(r_>/a;sa) , \label{modesum}
\end{eqnarray}
where $\Psi_n^{\,<}$ is determined by the boundary condition that it
be regular as $r\to 0$ and $\Psi_n^{\,>}$ by the condition that it
vanish at infinity.  Here we have introduced the standard notation
$r_<=\min(r,r')$ and $r_>=\max(r,r')$.  In contrast to the conical
space-time, the `boundary condition' as $r\to 0$ here is not an assumption
but is simply a consequence of the regularity of the space-time.  In
addition $\Psi_n^{\,<}$, $\Psi_n^{\,>}$ must satisfy the normalisation
condition
\begin{equation}
  {\partial \Psi_n^{\,<}(x;sa) \over \partial x}\Psi_n^{\,>}(x;sa) -
  \Psi_n^{\,<}(x;sa) {\partial \Psi_n^{\,>}(x;sa) \over \partial x} =
  {P(x) \over x} .  \label{normalisation}
\end{equation}

The Green function to Laplace's equation on the constant $t$ sections
may be found in an entirely analogous way. We find
\begin{eqnarray}
 G^{(3)}({\bf x},{\bf x'}) &=& \int\limits_{-\infty}^\infty \> {{\rm
   d}k \over 2\pi} e^{i k \Delta z} \sum_{n=-\infty}^{\infty}{\kappa
   \over 2 \pi } \> e^{in\kappa \Delta \phi} \Psi_n^{\,<}(r_</a;sa)
   \Psi_n^{\,>}(r_>/a;sa) ,\\ &=& {\kappa \over 2 \pi^2}
   \int\limits_0^\infty {\rm d}s \> \cos s\Delta z
   \sum_{n=-\infty}^{\infty} e^{in\kappa \Delta \phi}
   \Psi_n^{\,<}(r_</a;sa) \Psi_n^{\,>}(r_>/a;sa) , \label{3modesum}
\end{eqnarray}
where now $s \equiv |k|$ but all other symbols retain their previous
meaning.

As mentioned above, the `inner' mode function $\Psi^<$ is defined by
the boundary condition that it be regular as $r \to 0$.  As one
integrates out in the region $r<a$, it is impossible to write an
explicit formula for $\Psi^<$ without specifying the potential.  We
denote the solution to (\ref{radialeqn}) in this region by
$\Upsilon_n(r/a;sa)$.  However as $r$ increases beyond $a$, the
potential `turns off' and $\Psi^<$ becomes a sum of Bessel functions.
Thus we can write
\begin{equation}
\Psi_n^{\,<}(r/a;sa) = \cases{\Upsilon_n(r/a;sa) & for $r<a$, \cr &\cr
  A_n(sa) I_{\kappa |n|}(sr) + B_n(sa) K_{\kappa |n|}(sr) & for $r>a$. \cr}
  \label{psi_<}
\end{equation}
Here $A_n(sa)$ and $B_n(sa)$ are constants (with respect to $r$)
determined by matching $\Psi_n^{\,<}$ and its derivative at $r=a$.

The solutions to (\ref{radialeqn}) for the `outer' mode functions are
determined by the condition that they fall off when $r \to \infty$.
In the region $r>a$ where the potential vanishes, these solutions are
again Bessel functions.  Together with the normalization condition
(\ref{normalisation}) this yields
\begin{equation}
\Psi_n^{\,>}(r/a;sa)={1 \over A_n(sa)} K_{\kappa |n|}(sr) \quad{\rm for} \quad
               r>a.  \label{psi_>}
\end{equation}
We shall not need $\Psi_n^{\,>}$ within the region $r<a$ where the
potential is non-zero, since  we shall not attempt to compute any 
physical quantities inside the string core.

We now restrict ourselves to the region outside the core where both
$r$ and $r'$ are greater than $a$.  Then defining $C_n \equiv
B_n/A_n$, the Green functions on the true cosmic string may be written
as
\begin{eqnarray}
 G^{(4)}(x,&&x') = G^{(4)}_{\rm reg}(x,x') +\nonumber\\
       &&{\kappa \over 4 \pi^2}
\int\limits_0^\infty \>
       s {\rm d}s \> J_0\left(s \sqrt{\Delta t^2 + \Delta z^2}\right)
\sum_{n=-\infty}^{\infty} e^{in\kappa \Delta \phi}
C_n(s a) K_{\kappa |n|}(sr)K_{\kappa |n|}(sr') , 
  \label{g}
\end{eqnarray}
and
\begin{eqnarray}
  G^{(3)}({\bf x},{\bf x'}) 
          = G^{(3)}_{\rm reg}(&&{\bf x},{\bf x'})
      +\nonumber\\
       &&  {\kappa \over 2 \pi^2}
\int\limits_0^\infty {\rm d}s  \cos s \Delta z 
\sum_{n=-\infty}^{\infty} e^{in\kappa \Delta \phi}
             C_n(s a) K_{\kappa |n|}(sr)K_{\kappa |n|}(sr').
  \label{g3}
\end{eqnarray}
Here $G^{(4)}_{\rm reg}(x,x')$ and 
$G^{(3)}_{\rm reg}({\bf x},{\bf x'})$ are the Green functions appropriate to
 the
idealized string space-time with regularity conditions placed at the
origin: 
\begin{equation}
   G^{(4)}_{\rm reg}(x,x') \equiv 
     {\kappa \over 4 \pi^2} \int\limits_0^\infty \>
       s {\rm d}s \> J_0\left(s \sqrt{\Delta t^2 + \Delta z^2}\right)
\sum_{n=-\infty}^{\infty} e^{in\kappa \Delta \phi}
I_{\kappa |n|}(sr_<)K_{\kappa |n|}(sr_>) , 
  \label{g4conemodesum}
\end{equation}
and
\begin{equation}
  G^{(3)}_{\rm reg}({\bf x},{\bf x'}) \equiv
{\kappa \over 2 \pi^2} \int\limits_0^\infty {\rm d}s  \cos {s \Delta z}
\sum_{n=-\infty}^{\infty} e^{in\kappa \Delta \phi}
             I_{\kappa |n|}(sr_<)K_{\kappa |n|}(sr_>) .
  \label{g3conemodesum}
\end{equation}
The only dependence of $G^{(4)}$ and $G^{(3)}$ 
upon $V$ and $a$ or indeed upon the detailed structure of the space here is 
through the ratios $C_n(sa)$.  

For completeness we note that in the case of the regular Green functions on
the idealized cone one may perform the mode sums.
The Green function for the `wave equation' is \cite{dowker}
\begin{equation}
G^{(4)}_{\rm reg}(x,x') = 
    {1 \over 8 \pi^2}  {\kappa \sinh \kappa\eta \over  rr' \sinh\eta
  (\cosh \kappa\eta - \cos \kappa \Delta\phi)}\quad , \label{gcone}
\end{equation}
where 
\begin{equation}
 \cosh\eta \equiv {\Delta t^2 + \Delta z^2+ r^2 +{r'}^2 \over 2rr'}\quad,
                           \label{eta}
\end{equation}
with $\Delta t = t-t'$ and likewise for $\phi$ and $z$.
The Green function for Laplace's equation on the spatial section 
is \cite{smith}
\begin{equation}
  G^{(3)}_{\rm reg}({\bf x},{\bf x'}) = 
{1 \over 4 \pi^2 (2rr')^{1 \over 2}} \int\limits_\zeta^\infty
    {{{\rm d}s} \over (\cosh s - \cosh \zeta)^{1 \over 2}}
      {\kappa \sinh \kappa s \over  
  (\cosh \kappa s  - \cos \kappa \Delta\phi)}\quad , \label{g3cone}
\end{equation}
where 
\begin{equation}
 \cosh\zeta \equiv {\Delta z^2+ r^2 +{r'}^2 \over 2rr'}\quad.
                           \label{zeta}
\end{equation}

\section{APPROXIMATION}
\label{approximation}

The limit as the dimensionless variable $sa$ tends to zero  may be
considered either as the limit as the size of the string tends to zero
for fixed energy (as in \cite{AO}) or as the limit as the
scattering energy tends to zero for fixed string size (as in
\cite{KS}).  We now consider the
behavior of $C_n(sa)$ in this limit. First we write $C_n(sa)$ in terms
 of the solution to the radial wave equation
(\ref{radialeqn}), using the continuity of $\Psi_n^{\,<}$ and its
 derivative at $r = a$:
\begin{equation}
   C_n(sa) = 
  - {\displaystyle 
 I_{\kappa |n|}(sa){\partial \Upsilon_n \over \partial x} (1;sa) - 
                  sa I'_{\kappa |n|}(sa)\Upsilon_n (1;sa) \over
   \displaystyle
          K_{\kappa |n|}(sa){\partial {\Upsilon_n} \over \partial x} (1;sa) - 
                    sa K'_{\kappa |n|}(sa)\Upsilon_n (1;sa)} . 
                                            \label{c1} 
\end{equation}
It is convenient to rewrite this equation in the form
\begin{equation}
   C_n(sa) = 
  - {\displaystyle {1 \over \Upsilon_n (1;sa)}
              {\partial \Upsilon_n \over \partial x} (1;sa) - 
        { sa I'_{\kappa |n|} (sa) \over I_{\kappa |n|}(sa)} \over 
     \displaystyle {1 \over \Upsilon_n (1;sa)} 
                   {\partial \Upsilon_n \over \partial x} (1;sa) - 
        { sa K'_{\kappa |n|} (sa) \over K_{\kappa |n|}(sa)} } 
                   { I_{\kappa |n|}(sa) \over K_{\kappa |n|}(sa)} .
                                        \label{c2}
\end{equation}

It follows that for $n \neq 0$
\begin{equation}
    C_n(sa) =  - {\alpha_n - {\kappa |n|}\over \alpha_n + {\kappa |n|}}
  {2 \over \Gamma({\kappa |n|})\Gamma({\kappa |n|}+1)} 
          \> \left({sa / 2}\right)^{2{\kappa |n|}} +
    O\left( (sa)^{2{\kappa |n|}+2} \right),
                    \label{asymp}
\end{equation}
as $sa \to 0$ where 
\begin{equation}
\alpha_n \equiv \displaystyle {1 \over \Upsilon_n (1;0)}
{\partial \Upsilon_n \over \partial x} (1;0).
\label{alphadef}
\end{equation}
It is easy to see, directly from (\ref{radialeqn}) that $\alpha_n > 0$
so the denominator in (\ref{asymp}) cannot vanish.  (To see this, note
that $\Upsilon_n'(0,0)=0$ while we may assume that $\Upsilon_n(0,0)$
is positive.  Eq. (\ref{radialeqn}) then ensures that
$\bigl(x/P(x)\bigr)\Upsilon'(x,0)$ and hence $\Upsilon'(x,0)$
increases whereupon $\Upsilon(x,0)$ must increase.)  Thus, for $n \ne
0$, $C_n(sa)$ vanishes at least as fast as $(sa)^{2{\kappa |n|}}$ as $sa \to
0$.  On the other hand, in general $C_0(sa)$ vanishes only as an
inverse logarithm in this limit:
\begin{equation}
   C_0(sa) = {\alpha_0 \over 
        \alpha_0 \left[\ln \left( {sa / 2} \right) + {\cal C} \right]
                      - 1} + O\left( \left({sa \over \ln(sa)}\right)^2 \right),
          \label{asymp0}
\end{equation}
as $sa \to 0$, where $\cal C$ is Euler's constant.

Later we shall find it useful to rewrite the long-range term in 
Eq.(\ref{asymp0}) in the form
\begin{equation}
   C_0(sa)_{\rm long-range} = {1 \over \ln \left( {s/ q} \right)} 
                            \label{asympq}
\end{equation}
where 
\begin{equation}
q=2e^{-{\cal C}}/R \label{qR}
\end{equation}
and where $R$ is defined, in turn, by
\begin{equation}
R=a\exp(-\alpha_0^{-1}) \label{prefit}
\end{equation}
(The reason why we write things in this way, and the significance of the
interrelated quantities $q$ and $R$ will become clear below.)

Here again, we can easily see, directly from (\ref{radialeqn}) that in
the case $n=0$, $\alpha_0 \geq 0$ with $\alpha_0 =0$ if and
only if the potential, $V$, vanishes identically.  This includes the
particular case of minimal coupling ($\xi=0$) with no other potential.
Thus, in this case there are no long-range effects and the theory on the
idealized cone with regularity conditions accurately models the full
theory.  In all other cases, it seems reasonable, in view of
(\ref{asymp}) and (\ref{asymp0}), to approximate, say, the Green
function  $G^{(3)}({\bf x},{\bf x'})$ far from the string 
by dropping all terms other than
the $n=0$ term in the sum in (\ref{g3}) and substituting for  $C_0(sa)$
the long-range approximation $C_0(sa)_{\rm long-range}$ given by 
 (\ref{asympq}). 
This leads to the formula
\begin{equation}
 \qquad G^{(3)}_R({\bf x},{\bf x'}) = G^{(3)}_{\rm reg}({\bf x},{\bf x'}) +
{\kappa\over 2\pi^2} \int\limits_0^\infty {\rm d}s \> \cos {s \Delta z}
{K_0(sr)K_0(sr')\over \ln\left({s/ q}\right)}.
  \label{gR}
\end{equation}

As it stands, the integral in Eq. (\ref{gR}) is ill-defined because of the
pole in the integrand at $s=q$.
In an attempt to resolve this issue (and because it is of independent
interest) we now  discuss how this approximate Green function
arises from the point of view of Kay and Studer.  In \cite{KS} it is
argued that the low-energy dynamics for the (un-Euclideanized)
equation
\begin{equation}
\left({\partial^2\over\partial t^2}- \triangle + {1 \over a^2}V(r/a)\right)
    \varphi_{\rm true}(t,{\bf x})=0 
                      \label{uneucltrue}
\end{equation}
on the true string should be well approximated by solving the equation
\begin{equation}
\left({\partial^2\over\partial t^2}- \triangle_R\right)\varphi_R(t,{\bf x})=0 
                      \label{uneuclR}
\end{equation}
on the idealized string, where $\triangle_R$ is chosen to be the
($z$-translationally invariant) self-adjoint extension of the
Laplacian (defined on the domain of smooth functions compactly
supported away from $z=0$ in the Hilbert space of square integrable
functions) on the constant $t$ spatial sections of the idealized
string which gives the best fit to the low energy ``true'' dynamics of
equation (\ref{uneucltrue}).  Before we explain how this best-fit
choice is made, we note first that this choice amounts to a choice of
self-adjoint extension $\triangle^{(2)}_R$ of the 2-dimensional
Laplacian $\triangle^{(2)}$ (defined on the domain of smooth functions
compactly supported away from the origin in the Hilbert space of
square integrable functions) on the 2-dimensional ideal cone of
constant $t$ and $z$ since, in a sense made precise in \cite{KS}, each
translationally invariant 3-dimensional self-adjoint extension
$\triangle_R$ must arise as ${\partial^2\over\partial
z^2}+\triangle^{(2)}_R$ for some choice $\triangle^{(2)}_R$ of
2-dimensional self-adjoint extension.  Each of these self-adjoint
extensions is, in turn, as we have anticipated by our notation, known
to be labelled by a single parameter $R$ and corresponds to the
boundary condition at small $r$ on the $n=0$ sector component of
$\varphi$ (i.e. on the circular average of $\varphi$),
\begin{equation}
\varphi^{n=0}_R(t,r,z) \sim {\rm c}(t)\ln(r/R), \label{kayasymp2}
\end{equation}
where the ``time-dependent constant'' ${\rm c}(t)$ is independent of
$z$, while regularity holds in all the sectors with $n\ne 0$.  So,
solving (\ref{uneuclR}) amounts to solving the equation $\displaystyle
({\partial^2\over\partial t^2}- \triangle_R)\varphi(t,{\bf x})=0$
subject to these boundary conditions.

Turning to the question of how the best fit is made, we begin by
remarking that, at {\it zero} energy, the ($z$-translationally
invariant, cylindrically symmetric) solution to (\ref{uneucltrue})
will take the exact form
\begin{equation}
\varphi^{\rm static}_{\rm true}(r) = {\rm const}{\cdot}\ln(r/R), 
                        \label{stattrue}
\end{equation}
outside the support of the potential for some positive real parameter $R$
which, it is worth noticing, will be related to the logarithmic derivative of
$\varphi^{\rm static}_{\rm true}$ at $r=a$ by
\begin{equation}
\left. { r \over \varphi^{\rm static}_{\rm true}} 
              {{\rm d} \varphi^{\rm static}_{\rm true} \over {\rm d} r} 
\right|_{r=a} =
                      {1 \over \ln(a/R)} ,  \label{logderiv}
\end{equation}
or equivalently,
\begin{equation}
   R = a \exp{\left(- {1 \left/ \left. 
{ r \over \varphi^{\rm static}_{\rm true}} 
              {{\rm d} \varphi^{\rm static}_{\rm true} \over {\rm d} r} 
\right|_{r=a}  \right. }\right)}  .  \label{fitting}
\end{equation}

The best fit self-adjoint extension is then
declared to be  the one for which the $R$ in (\ref{kayasymp2})
coincides with the
$R$ in (\ref{stattrue}):  In the language of \cite{KS}, one identifies

the label
of the self-adjoint extension in (\ref{uneuclR}) with the ``scattering length''
of the potential in (\ref{uneucltrue}) which may be calculated from the
``fitting formula'' (\ref{fitting}). 
 
We remark that, mathematically, there is
clearly no distinction between the zero energy solution $\varphi^{\rm
static}_{\rm true}(r)$ to (\ref{uneucltrue}) in the region $r \le a$ 
and the zero
$sa$ regular solution $\Upsilon_0(r/a;0)$ to (\ref{radialeqn}) so that 
the quantity $R$ introduced above in (\ref{prefit}) is now seen to be 
identical with Kay and Studer's scattering length and the equation
(\ref{prefit}) to be mathematically identical with the fitting formula
(\ref{fitting}).   Note that as $\alpha_0 \geq 0$, one can read off from
(\ref{prefit}) that the scattering length $R$ is always bounded by the
 string size
$a$.  (This is the content of ``Observation 1'' in Section 5.2 of \cite{KS}.)

The significance of $q$, related to $R$ by Eq.(\ref{qR}), is also
explained by Kay and Studer: $-q^2$ is the eigenvalue -- with
normalizable eigenstate which we call $\psi^{(q)}_{\rm bound}$
($\psi^{(q)}_{\rm bound}(r)=\pi^{-{1\over 2}}qK_0(qr)$, see equation
(2.6) of \cite{KS}) -- which exists for (minus) the self-adjoint
extension of the 2-dimensional Laplacian $-\triangle^{(2)}_R$ for $R
\neq 0$. If one thinks of $-\triangle^{(2)}_R$ as a possible candidate
for a Schr\"odinger operator for quantum mechanics on the cone, then
$\psi^{(q)}_{\rm bound}$ would have a physical interpretation as a
``bound state''. As is appropriate for a bound state, this eigenvalue
of minus the Laplacian is {\it negative}.  Of course, since we are
assuming our potentials to be non-negative, the corresponding ``true
2-dimensional Laplacians'' for the true smooth string space-times
considered here can have {\it no} bound states, so the bound state in
the idealized approximation is a mathematical artefact. (In the
language of \cite{KS} and \cite{KF} it is related to small scale
aspects of the idealized dynamics and hence not ``believable''.  See
``Pitfall 3'' in Note 20 in \cite{KS}.)

The relevant solution for $R=0$ is simply $\varphi_0 = {\rm const}$
and so $\alpha_0 =0$. In this case $q$ is set by convention to zero
(see Note 4 in \cite{KS}), and the corresponding extension
$-\triangle^{(2)}_0$ (which is actually the Friedrichs extension) of
(minus) the 2-dimensional cone Laplacian does not possess a bound
state.

The circle of ideas may now be completed since one may show (for
example by an elegant method using the Krein resolvent formula -- See
Appendix \ref{krein}) that the {\it exact} Green function for the
approximate Green function equation
\begin{equation}
     - \triangle_R G^{(3)}_R({\bf x},{\bf x'}) = 
                        \delta^{(3)}({\bf x},{\bf x'}), 
                          \label{3Reqn}
\end{equation}
on the idealized string is identical with the approximate
expression (\ref{gR}) given earlier for the exact 3-dimensional Green function
outside $r=a$ on the (constant $t$ sections of the) true string. 
Moreover, as we mention in Appendix \ref{krein}, the problem of the
pole at $s=q$ in (\ref{gR}) should be resolved by a principal part
prescription.

Clearly, there will be a similar formula to (\ref{gR}) for an
idealized 4-dimensional Green function $G^{(4)}_R(x,x')$ for each
$R$-value:
\begin{equation}
 G^{(4)}_R(x,x') = G^{(4)}_{\rm reg}(x,x') + 
{\kappa\over 4\pi^2} \int\limits_0^\infty s{\rm d}s \> 
    J_0\left(s\sqrt{\Delta t^2 + \Delta z^2} \right)
      {K_0(sr)K_0(sr')\over \ln\left({s/q}\right)}.
  \label{gR4s}
\end{equation}
which again should be understood as a principal part integral.
As before, one may arrive at this by either of two routes:
 the first by dropping
all but the zero term in the sum in (\ref{g}) and replacing $C_0$ there by the
approximation (\ref{asymp}), the second by calculating directly, e.g. by the
Krein resolvent formula method of Appendix \ref{krein},
 the Green function for the ideal operator
${\partial^2\over\partial t^2}+{\partial^2\over\partial z^2}
+\triangle^{(2)}_R$.

We remark that, in both (\ref{gR}) and (\ref{gR4s}), the integrands have
poles at $s=q$ related to the existence of the ``bound state'' in
$- \triangle^{(2)}_R$ that we mentioned above.  In view of this structure,
and in contrast with the case of the true 4-dimensional Green
function $G^{(4)}(x,x')$ of Eq. (\ref{g}), we do {\it not} expect  the
idealized Green function $G^{(4)}_R(x,x')$ of Eq.(\ref{gR4s}) to exactly
correspond under analytic continuation with any exact two-point function
for the quantum field theory in the Lorentzian version of the idealized
string spacetime in some ground state,  (i.e. we expect the appropriate
Osterwalder-Schrader-like axioms to fail for $G^{(4)}_R(x,x')$.)  In
fact, as was discussed in \cite{KS}, (except for the case $R=0$) the
field algebra for the Lorentzian idealized string will not admit a
ground state for the time-evolution corresponding to the classical
solutions to the massless Klein-Gordon equation with the boundary
conditions (\ref{kayasymp2}). This is a physically spurious result which
has to do with the ``unbelievability'' of the bound state, as is
discussed in some detail in \cite{KS} (see also \cite{KF}).
In the last paragraph of Section
5 there, one possible method of circumventing this problem is proposed
which involves ``projecting out'' the ``bound state contribution'' to
the exact dynamics for the boundary condition (\ref{kayasymp2}) to
obtain another dynamics which approximates it on large scales and admits
a quantum ground state, but which is ``non-local on small scales''.  It
is then proposed in \cite{KS} to study the quantity 
${\langle {\hat \varphi}^2 \rangle}$ in this
ground state in order to compare with the results of \cite{AO}.  In the
event, we have performed such a study in the present paper (see next
section) but we circumvent the ``unbelievability problem'' in an, at
least superficially, somewhat different way to that proposed in
\cite{KS}, namely,  by essentially eliminating in a rather  direct way (which
we explain in the next section) the pole in (\ref{gR4s}).  It might be
interesting to investigate further the relationship between the approach
adopted here and that proposed in the final paragraph of Section 5 of
\cite{KS}.   

Next we consider the question of determining the scattering length of a given
potential term $V$.  A case of particular interest is that of weak potentials. 
For $s=n=0$, Eq.(\ref{radialeqn}) reduces to,  
\begin{equation}
\biggl[- {1 \over xP(x)}{{\rm d} \over {\rm d}x}
{x \over P(x)} {{\rm d} \over {\rm d}x}  + V(x)
      \biggr] \Psi_0(x;0) =  0 .     
                             \label{0eqn}
\end{equation}
For weak potentials it follows that $\Upsilon(1;0) \approx \Upsilon(0;0)$
and
\[
 {\partial \Upsilon \over \partial x} (1;0) \approx
    \int\limits_0^1 {\rm d}x \>  x P(x) V(x) \Upsilon(0;0)
\]
so
\begin{equation}
   R \approx a \exp \left( - 1 \left/ \int\limits_0^1 {\rm d}x \>  x P(x) V(x) 
                \right. \right) .  \label{weak}
\end{equation}   
This is  exactly Eq.\ (5.10) of Ref.\ \cite{KS} written in our current
conventions. We may rewrite Eq.\ (\ref{weak}) as 
\begin{equation}
   R \approx a \exp \left( - 2 \pi \left/ 
        \kappa \int \sqrt{ {}^{(2)}g }{\rm d}^2x \> V(x) \right.
                \right) .  \label{covweak}
\end{equation}   
It is remarkable that for the standard curvature coupling potential,
$V = \xi {\cal R}$, the integral in Eq.\ (\ref{covweak}) is a 
topological invariant given by 
\begin{equation}
\int \sqrt{ {}^{(2)}g }\,{\rm d}^2x \> {\cal R}(x) =
       4\pi (1 - 1/\kappa) ,  \label{topinv}
\end{equation}
and correspondingly 
\begin{equation}
   R \approx a \exp \left( - {1 \over  
       2 \xi  (\kappa - 1) }
                \right) .  \label{Rtopinv}
\end{equation} 
For $\xi=1/6$ and a GUT scale string, $\kappa = 1 + 10^{-6}$, we have
\begin{equation}
   R \approx e^{-3,000,000} a .  \label{RGUT} \end{equation} 
We will show in the next section that despite such an incredibly small
size for $R$, it will give rise to large relative corrections to 
${\langle {\hat \varphi}^2 \rangle}$
and ${\langle {\hat T}_{\mu}{}^{\nu} \rangle}$ on cosmological scales.

As a useful check we calculate $R$ for the `flower-pot' and `ballpoint
pen' models of Allen \& Ottewill \cite{AO}. A little care is required
for the `flower-pot' model as the `inner' mode function $\Psi_<$ has a
first derivative which is not continuous at $r=a$, as is apparent from
Eq. (17) of Ref. \cite{AO}.  It is important in this case that the
value of $\alpha_0$, which appears in the definition of $R$, be
evaluated by taking the right derivative of $\Psi_<$ at $r=a+0$.  By
this means, the appropriate $\alpha_0$ may be obtained from a
comparison with Eq.(\ref{alphadef}).  For the `flower-pot' model we
find\cite{AOcorr}
\begin{equation}
        R_F = a \exp \left( - {1 \over 2 \xi (\kappa -1)} \right) , 
                   \label{R_F}
\end{equation}
exactly. For the `ballpoint pen' model we find
\begin{equation}
        R_B =  a \exp 
     \left(  {\kappa  P_{\nu_0}(1/\kappa) \over (\kappa^2 -1)
                   P_{\nu_0}'(1/\kappa)} \right) , \end{equation} where
$P_{\nu_0}$ denotes the Legendre function of the first kind, and
$\nu_0(\nu_0+1)  \equiv -2\xi $. For \hbox{$(\kappa-1) <\!< 1$}, this reduces
to Eq.\ (\ref{Rtopinv}) on noting that 
$ P_{\nu_0}(1/\kappa) \approx  P_{\nu_0
}(1) =1$ and $ P_{\nu_0}'(1/\kappa) \approx  P_{\nu_0}'(1) = - \xi$.

\section{VACUUM EXPECTATION VALUES}
\label{expectation} 

In this section we shall investigate to what extent the expectation
values 
of ${\langle {\hat \varphi}^2 \rangle}$ and
${\langle {\hat T}_{\mu}{}^{\nu} \rangle}$ on rounded cones for
 non-minimally coupled
fields can be mocked up by choosing 
the  appropriate non-zero $R$ value. We
start by considering the renormalised  expectation value of 
${\hat\varphi}^2$.   This may be defined as \cite{K,BO} 
\begin{equation}
 \langle \hat \varphi^2(x) \rangle 
 = \lim_{x' \to x} \left[G^{(4)}(x,x')- G^{(4)}_{\rm Euclidean}(x,x')\right] 
                \label{renorm},
\end{equation}
where $G^{(4)}_{\rm Euclidean}(x,x')={1/\bigl( 8 \pi^2 \sigma(x,x')\bigr)}$ 
is the Green function for flat 4-dimensional Euclidean space. 
(Here, $2 \sigma(x,x')$ denotes the square of the geodesic distance
from $x$ to $x'$.) By symmetry ${\langle {\hat \varphi}^2 \rangle}$ is
a function only of $r$.  From Eq.(\ref{g}) we may write
\begin{eqnarray}
  {\langle {\hat \varphi}^2 \rangle} &=& 
           {\langle {\hat \varphi}^2 \rangle}_{\rm reg} +
       {\kappa \over 4\pi^2} \int\limits_0^\infty s{\rm d}s
\sum_{n=-\infty}^{\infty} C_n(s a) K_{\kappa |n|}^2(sr) \nonumber\\
  &=& {\langle {\hat \varphi}^2 \rangle}_{\rm reg} +
       {\kappa \over 4\pi^2 r^2} \int\limits_0^\infty v{\rm d}v
\sum_{n=-\infty}^{\infty} C_n(v {a \over r}) K_{\kappa |n|}^2(v) 
                                                   \label{phi2} 
\end{eqnarray}
where the first term is the renormalised vacuum expectation value of 
${\hat \varphi}^2$ on the idealized cone with regularity conditions
imposed at the string \cite{dowker}.  

Bearing in mind the asymptotic behavior of the $C_n$ for small
argument, one might hope to approximate ${\langle {\hat \varphi}^2
  \rangle}$ 
for $r > a$ by
neglecting the terms in Eq.(\ref{phi2}) corresponding to $n \neq 0$
and replacing $ C_0(v {a \over r})$ by its asymptotic
form for small argument given by Eq.(\ref{asympq}).  This is of course
equivalent to replacing $G^{(4)}(x,x')$ in (\ref{renorm}) by the
approximate Green function $G_R^{(4)}(x,x')$ of (\ref{gR4s}).
Whichever of these points of view one adopts, the correction term in
Eq.(\ref{phi2}) would then be approximated by 
\begin{equation} 
{\kappa
\over 4\pi^2 r^2} \int\limits_0^\infty v{\rm d}v\> {K_0^{\,2}(v) \over
\ln v - \ln (qr) }.  \label{correction}
\end{equation}
While one approach would be to stop at this point, interpreting
(\ref{correction}) as a principal part integral (cf. the discussion in
Sec.~\ref{approximation}) we shall now argue for a simpler
approximation which we have reason to expect to be no less accurate.
In fact, as is clear from the discussion of Sec.~\ref{approximation},
the pole at $v=qr$ in the integrand of (\ref{correction}) lies well
beyond the range of $v\  (=rs)$ for which the approximation
(\ref{asympq}) has any validity. 
If we return to the exact $n=0$ term,
\begin{equation}
       {\kappa \over 4\pi^2 r^2} \int\limits_0^\infty v{\rm d}v \>
              C_0(v {a \over r}) K_0^{\,2}(v),           \label{0term}
\end{equation}
we see that the pole occurs because we are  using the small $sa$ asymptotic
form for $C_0(sa)$ when $sa=2e^{ - {\cal C}}e^{1 /\alpha_0}$.  However
this is always greater (and generally much greater) than one.  In fact
$C_0(sa)$ cannot have any singularities, as these would correspond to
zeros in $A_0(sa)$ of (\ref{psi_<}).  These may be ruled out by
recalling from Section \ref{approximation} that $\alpha_0$ (which must
match onto the logarithmic derivative of $\Psi_n^{<}(r/a;sa)$ of
(\ref{psi_<})) is necessarily positive, while the logarithmic
derivative of the Bessel function $K_0(sr)$ is negative.  (This
absence of zeros in $A_0(sa)$ corresponds to the absence of any bound
state in [the $n=0$ sector of] the differential operator in
(\ref{potentialeqn}) when regarded as a Schr\"odinger operator.  See
the end of Section 5 of Kay and Studer \cite{KS} as discussed in the
previous Section.)

\begin{figure}
\centerline{\epsfig{figure=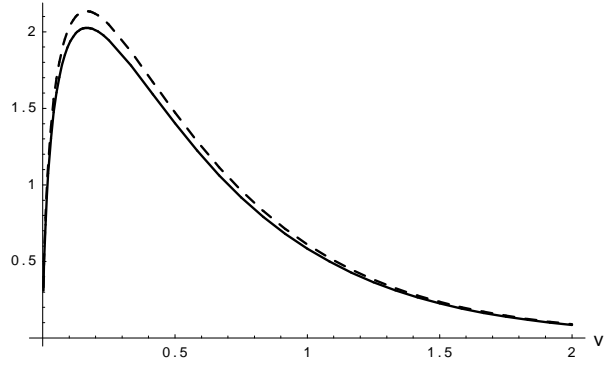,height=9truecm,
      bbllx=3truecm,bblly=1.5truecm
      ,bburx=17truecm,bbury=23.5truecm, angle=270} }
\caption{ The solid curve is ($- 10^3$ times) the integrand 
of the exact $n=0$ term, given by (\ref{0term}), and the dashed curve
is ($- 10^3$ times) the approximate integrand, given by 
(\ref{approxintegrand}) including the prefactor,   for
the flower-pot model with $\kappa= 100/99$ and $r/a= 10^{3}$.
The agreement between the two curves increases as $\kappa$ gets closer
to $1$ or as $r/a$ is increased.}
\label{integrandplot}
\end{figure}

This observation suggests that an equally satisfactory 
approximation will be given by simply replacing the integral 
\begin{equation}
\int\limits_0^\infty    v{\rm d}v\>
         {K_0^{\,2}(v) \over \ln v - \ln (qr) }
\end{equation}
in (\ref{correction}) by
\begin{equation}
\label{approxintegrand}
-{1\over \ln qr}\int\limits_0^\infty    v{\rm d}v\>
         K_0^{\,2}(v) 
\end{equation}
which, by the identity given in Appendix \ref{identity}  is equal to 
\begin{equation}
-{1\over 2\ln (qr)}.
\end{equation}
The rationale behind this is as follows: The multiplier $vK_0^2(v)$ in
the integrand of (\ref{0term}) vanishes at $v=0$, peaks around $v$ of
order $1$ and decays as $\exp (-2v)$ for large $v$.  On the other
hand, $C_0(v{a \over r})$ vanishes at $v=0$ and grows less slowly than
$\exp(2v {a \over r})$ for large $v$. Thus, assuming that $C_0(v{a
\over r})$ is sufficiently well behaved, the contribution to the
integral from the ``large $v$'' region (where ${1/ \ln (v/qr)}$ no
longer well approximates $C_0(v{a \over r})$ and where its
pole is located) will be negligible.  In
addition, we are interested in regions far from the string so that
$\ln(qr)$ (which is always greater than $\ln(r/a)$) will be large.
Thus, except when $v$ is very small, where again one expects the
contribution to the integral to be small, ${1 / \ln (v/qr)}$ will not
only well approximate $C_0(v{a \over r})$, but will also be slowly
varying and, in its turn, well approximated by ${-1 / \ln (qr)}$.  The
exact integrand and its approxiamtion are illustrated for a 
 `flower-pot' model string in
Fig.~\ref{integrandplot}.

In conclusion, we have the approximation
\begin{equation}
 {\langle {\hat \varphi}^2 \rangle}_{\rm R}      =
      {\kappa^2 - 1 \over 48\pi^2 r^2} 
             - {\kappa \over 8\pi^2 r^2 \ln (qr)} =
 {\kappa^2 - 1 \over 48\pi^2 r^2} 
             - {\kappa \over 8\pi^2 r^2 \ln (2e^{-{\cal C}}r/R)}
 ,    \label{approxphi2}
\end{equation}
where the first term is simply 
${\langle {\hat \varphi}^2 \rangle}_{\rm reg}$ -- i.e. the
value one would obtain on the assumption of an ideal string with
regular boundary conditions\cite{smith}. We see here directly the long-range
effect of the cosmic string structure, parameterised by the single
parameter $q$ or equivalently $R$.

\begin{figure}
\centerline{\epsfig{figure=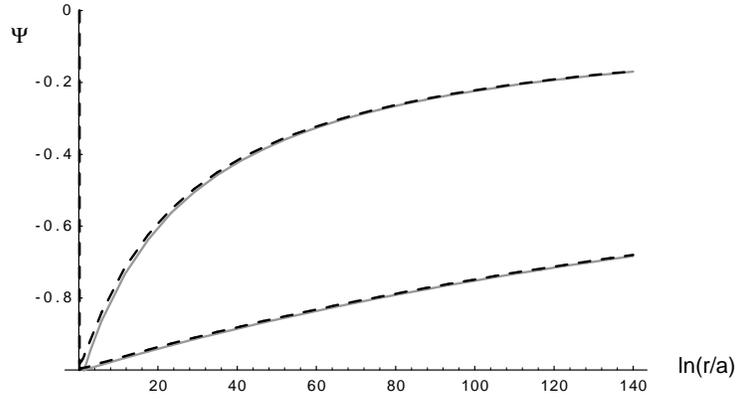,height=9truecm,
      bbllx=1.5truecm,bblly=2.5truecm
      ,bburx=16truecm,bbury=22truecm,angle=270} }
\caption{The solid curves are the exact relative correction $\Psi(r/a)$,
  given by Eq.~(\ref{psi}), and the dashed curves are   
approximate relative correction,
given by Eq.~(\ref{psiR}),
 for the flower-pot model with $\xi = 1/6$ and $\kappa = 10/9$
(upper curves) and $\kappa = 100/99$ (lower curves).}
\label{phi2plot}
\end{figure}

While our above argument for the approximation (\ref{approxphi2}) is
not justified by any rigorous bound, we believe it is likely to be an
excellent approximation in practice whenever $r$ is much greater than
the string radius.  Evidence for this may be seen immediately in
Fig.~\ref{phi2plot} where we plot the exact expression for
\begin{equation} 
\label{psi}
\Psi (r) \equiv {{\langle {\hat \varphi}^2 \rangle} - 
{\langle {\hat \varphi}^2 \rangle}_{\rm reg} \over 
{\langle {\hat \varphi}^2 \rangle}_{\rm reg} }
\end{equation}
for a `flower-pot' model with $\xi = {1 \over 6}$ against its
approximation
\begin{equation}
\label{psiR}
  \Psi_R(r) = - {6 \kappa \over \kappa^2 -1} {1 \over \ln (qr)} = 
       - {6 \kappa \over \kappa^2 -1} 
          {1 \over  \ln (2e^{-{\cal C}}r/R)} 
\end{equation}
using $R_F$ as given in Eq.(\ref{R_F}).
The calculation of the exact curve was a substantial computational chore 
while the calculation of the approximation curve is clearly trivial.

To determine the importance of the correction term for a GUT string we
notice that for such a string $(\kappa-1) <\!< 1$ so we may make the
weak potential approximation (\ref{Rtopinv}) whereupon
(\ref{approxphi2}) becomes
\begin{equation}
 {\langle {\hat \varphi}^2 \rangle}_{\rm R}      =
      {\kappa - 1 \over 24\pi^2 r^2}\left[1 -
       {6\xi \over 2\xi(\kappa -1) \ln (2e^{-C}r / a) + 1 }  \right ]. 
                   \label{approxphi3}
\end{equation}
whereupon we see that, for typical non-zero values of $\xi$ the
correction term is of the same order of magnitude as the first term
for all reasonable large values of $r$ and vanishes so slowly as $r
\to \infty$, that one needs to consider $r$ values which massively
exceed the radius of the observable universe before the correction
term is significantly attenuated!  For example, in the most
interesting case of conformal coupling, $\xi=1/6$, (which,
incidentally, is special in that, for this value of $\xi$, the
correction term almost precisely cancels the first term at reasonable
values of $r$) one requires
\begin{equation}
r ={a \over 2e^{-C}}e^{3/(\kappa-1)} \approx e^{3,000,000} {\rm cm} 
\end{equation}
before ${\langle {\hat \varphi}^2 \rangle}$ has climbed back up to
 half its asymptotic value of
 ${\langle {\hat \varphi}^2 \rangle}_{\rm reg}$.

Given the success of the above approximation scheme for 
${\langle {\hat \varphi}^2 \rangle}_R$ it
is tempting to extend it to ${\langle {\hat T}_{\mu}{}^{\nu}
  \rangle}$, the renormalised vacuum 
expectation value of the stress tensor.  Starting by keeping only the $n=0$ 
term in (\ref{g}) we may write
\begin{eqnarray}
  {\langle {\hat T}_{t}{}^{t} \rangle} &=& 
         {\langle {\hat T}_{t}{}^{t} \rangle}_{\rm reg} +
    {\kappa \over 4 \pi^2}\int\limits_0^\infty s^3 {\rm d}s \>
   \left\{ 2\xi K_0^{\,2}(sr) + (2\xi-{\textstyle {1 \over 2}})K_1^{\,2}(sr)
                                   \right\} C_0(sa) \label{Ttt}\\
  {\langle {\hat T}_{r}{}^{r} \rangle} &=&
            {\langle {\hat T}_{r}{}^{r} \rangle}_{\rm reg} +
        {\kappa \over 4 \pi^2}\int\limits_0^\infty s^3 {\rm d}s \>
   \left\{ -{\textstyle {1 \over 2}} K_0^{\,2}(sr) - 
          {2\xi \over sr}K_0(sr)K_1(sr)
    + {\textstyle {1 \over 2}} K_1^{\,2}(sr)   \right\} C_0(sa) \label{Trr}\\
  {\langle {\hat T}_{\phi}{}^{\phi} \rangle} 
  &=& {\langle {\hat T}_{\phi}{}^{\phi} \rangle}_{\rm reg} + \nonumber\\
&&{\kappa \over 4 \pi^2}\int\limits_0^\infty s^3 {\rm d}s \>
   \left\{(2\xi -{\textstyle {1 \over 2}}) K_0^{\,2}(sr) + 
{2\xi \over sr}K_0(sr)K_1(sr)
    + (2\xi - {\textstyle {1 \over 2}})K_1^{\,2}(sr)   \right\} 
                        C_0(sa)\quad         \label{Tphiphi}
\end{eqnarray}
and, by boost invariance in the $t$-$z$ plane, 
${\langle {\hat T}_{z}{}^{z} \rangle} = {\langle {\hat T}_{t}{}^{t} \rangle}$. 
Here ${\langle {\hat T}_{\mu}{}^{\nu} \rangle}_{\rm reg}$ is the
 standard result for
the idealized cone \cite{smith}:
\begin{equation}
{\langle {\hat T}_{\mu}{}^{\nu} \rangle}_{\rm reg}=  
   {\kappa^4 -1 \over 1440\pi^2 r^4}
      {\rm diag} (1,1,1,-3) + 
  {(\kappa^2 -1) \over 24\pi^2 r^4 }  (\xi - {\textstyle {1 \over 6}})
      {\rm diag}(2,2,-1,3)_\mu{}^\nu .
\end{equation}
  It is readily verified that 
these expressions satisfy the only non-trivial conservation equation
\begin{equation}
  {\rm d \, \over {\rm d}r } \left( r {\langle {\hat T}_{r}{}^{r}
      \rangle}
                  \right)
                 =  {\langle {\hat T}_{\phi}{}^{\phi}
                     \rangle}
                         . \label{conservation}
\end{equation}
We were guaranteed conservation here as the correction term we have kept to 
$G_{\rm reg}$ is a homogeneous solution to the wave equation.

    If we pursue the same line of argument as above then we are led to 
the approximation
\begin{equation}
{\langle {\hat T}_{\mu}{}^{\nu} \rangle}_R =
        {\langle {\hat T}_{\mu}{}^{\nu} \rangle} _{\rm reg}
   - {\kappa \over 4\pi^2 r^4 \ln (qr)}  (\xi - {\textstyle {1 \over 6}})
      {\rm diag}(2,2,-1,3)_\mu{}^\nu  , \label{approxTmunu}
\end{equation}
where we have made frequent use of the identity given in Appendix 
\ref{identity}.
In making the transition to the last expression we have moved away from an 
exact solution to the wave equation.  As a consequence it is not surprising 
that ${\langle {\hat T}_{\mu}{}^{\nu} \rangle}_{\rm R}$ 
violates conservation by terms of order
\begin{equation}
       {1 \over r^4 \left( \ln(qr)\right)^2 }  .
\end{equation}
Nevertheless, since we are interested in regions very far from the
string $\ln(qr)$ (which is always greater than $\ln(r/a)$) will be
large so that the violation is small and ${\langle {\hat
    T}_{\mu}{}^{\nu}  \rangle}_{\rm R}$ should
provide us with an acceptable approximation to the true stress tensor.
In particular, it is reasonable to conclude that for $0\le\xi\ne1/6$
the energy-density will have long-range corrections arising from the
string structure.  On the other hand, in distinction to the situation
for ${\langle {\hat \varphi}^2 \rangle}$, in the conformally coupled
 case, $\xi=1/6$, our
correction term for the stress tensor vanishes.

\section{SELF-FORCE}
\label{selfforce}

Similar techniques may be used to investigate the 
electrostatic self-force on a point test charge 
outside a superconducting cosmic string of finite thickness  \cite{KS}. 
Working in SI units, 
the electrostatic potential $\varphi({\bf x})$ due to a point 
charge $Q$ at ${\bf x}_0$ will be 
\begin{equation}
   \varphi({\bf x}) = {Q \over \epsilon_0} G^{(3)}({\bf x},{\bf x}_0) ,
                               \label{potential}
\end{equation} 
where $G^{(3)}({\bf x},{\bf x}_0)$ solves (cf. Eq.(\ref{3potentialeqn}))
\begin{equation}
\left(- \triangle + {1 \over a^2}V(r/a)\right)G^{(3)}({\bf x},{\bf x'}) = 
                        \delta^{(3)}({\bf x},{\bf x'}), 
\end{equation}
Here, $V$ represents the local photon mass term supported inside the
string radius which will be responsible for making the string
superconducting.  The detailed shape of $V$ will depend on the
particular model field theory out of which the string is made (see
\cite{witten}), but we shall assume  
 it to be non-negative.  $\triangle$ represents the usual
Laplace-Beltrami operator on scalars.  (This is the correct operator
here even though $\varphi$ is a component of a 4-vector, because of
the ultrastatic nature of the metric (\ref{ds2}).)

The renormalised self-energy $W({\bf x}_0)$, for a point ${\bf x}_0$
outside the string core, will be given by the formula
\begin{equation}
W({\bf x}_0)=\lim_{{\bf x}\rightarrow {\bf x}_0}{Q^2\over
    2\epsilon_0}(G^{(3)}({\bf x},{\bf x}_0)-
               G^{(3)}_{\rm Euclidean}({\bf x},{\bf x}_0))
                                                        \label{pointsplitU}
\end{equation}
where $G^{(3)}_{\rm Euclidean}({\bf x},{\bf x}_0) = 1/\bigl(4\pi  |{\bf x}
- {\bf x}_0|)$ 
is the corresponding Green function
one would have in the case the string were absent (i.e. if, in
Eq. (\ref{3potentialeqn}), $V$ were equal to zero, and $\triangle$ were the
usual flat space  Laplacian).  The self-force ${\bf F}$ is then given in
terms of the self-potential by
\begin{equation}
   {\bf F} = - \nabla W .
\end{equation}
Using Eq. (\ref{g3}), we obtain from Eq. (\ref{pointsplitU})
\begin{equation}
  W = W_{\rm reg} + W_{\rm super}
             \label{selfenergy}
\end{equation}  
where
\begin{equation}
W_{\rm super}={\kappa Q^2\over 4\pi^2\epsilon_0} \int\limits_0^\infty {\rm d}s
\sum_{n=-\infty}^{\infty} C_n(s a) K_{\kappa |n|}^2(sr) 
                                                   \label{Usuper} 
\end{equation}               
and
\begin{equation}
W_{\rm reg}({\bf x}_0)=\lim_{{\bf x}\rightarrow {\bf x}_0}{Q^2\over
    2\epsilon_0}(G^{(3)}_{\rm reg}({\bf x},{\bf x}_0)
         -G^{(3)}_{\rm Euclidean}({\bf x},{\bf x}_0 ))
                                                     \label{conenergy}
\end{equation}
is the renormalised self-energy appropriate to the 
idealized string with regularity conditions imposed at the string.  This
latter quantity, which may be regarded as the contribution to the
self-energy due to space-time curvature was first calculated by 
Smith \cite{smith} and
Linet \cite{linet}. 
This is the only contribution in the case of a non-superconducting
string and in this case there are no long range effects of the string
structure.  Now, combining 
(\ref{conenergy}) and (\ref{g3cone}), one easily obtains
\begin{equation}
  W_{\rm reg} (r)= { 1 \over 4\pi \epsilon_0}{ Q^2 {\cal K}(\kappa) \over
                     2 r} \label{Ucone}
\end{equation}
with
\begin{equation}
   {\cal K}(\kappa) \equiv  {1 \over  \pi} \int\limits_0^\infty {\rm d}v
      \> {\kappa \coth (\kappa v) - \coth v \over \sinh v} .   \label{calK}
\end{equation}
As shown in \cite{smith}, for $\kappa -1 <\!< 1$, ${\cal K}(\kappa) 
\approx  (\kappa -1)\pi /8$, so that $W_{\rm reg}$ corresponds to a
(repulsive) contribution
\begin{equation}
{\bf F}_{\rm reg} \approx  { Q^2(\kappa-1)  \over 64 \epsilon_0  }
                     {\hat {\bf r} \over r^2} \label{F0}
\end{equation}
to the self-force.  

We now turn to $W_{\rm super}$ which, as we shall see, turns out to be
attractive, and typically, very much larger in magnitude than $W_{\rm
reg}$: Following a similar path to that adopted in Section
\ref{expectation}, if one naively approximates (\ref{Usuper}) by
discarding all terms in the sum other than $n=0$, and replacing
$C_0(sa)$ by its asymptotic form (\ref{asympq}) one obtains the
formula
\begin{equation}
W_{\rm super}(r)= {Q^2 \kappa \over 4\pi^2\epsilon_0}
                       \int\limits_0^\infty {\rm d}s \>
                      {K_0^2(sr) \over \ln (s/q)}  
= {Q^2 \kappa \over 4\pi^2\epsilon_0 r}
                       \int\limits_0^\infty {\rm d}v  \>
                      {K_0^2(v) \over \ln v - \ln(qr)}    .
      \label{Uintapprov}  
\end{equation}
(Alternatively, one may obtain (\ref{Uintapprov}) by setting $\Delta
z=0$ and $r=r'$ in (\ref{gR}).)  Here, we recall (see around and after
Eq.(\ref{qR})) that $q=2{\rm e}^{-C}/R$ and $R$ is the scattering
length appropriate to Eq.(\ref{uneucltrue}) in the case where $V$
represents the local photon mass term.  As explained in \cite{KS} on
the basis of arguments given in \cite{witten}, one expects $R$ to be
of the order of the string radius $a$.  (It will certainly be bounded
by $a$ if, as we have assumed, $V$
is non-negative.)  Hence, $q$ in (\ref{Uintapprov}) will be of the
order of $1/a$.  As in the discussion of equation (\ref{correction})
the formula, (\ref{Uintapprov}) must be interpreted as a principal part
integral because of the pole at $v=qr$.  However, 
for similar reasons to those discussed in the case of 
(\ref{correction}), we expect that a simpler and still good
approximation will be given by replacing Eq. (\ref{Uintapprov}) with
\begin{equation}
W_{\rm R}(r) = - {Q^2 \kappa \over 4\pi^2\epsilon_0 r\ln(qr)}
                       \int\limits_0^\infty {\rm d}v
                       K_0^2(v) = -{\kappa Q^2 \over 16\epsilon_0 r\ln(qr)} 
           = -{\kappa Q^2 \over 16\epsilon_0 r\ln(2e^{-{\cal C}}r/R)},
     \label{Uintimprov}  
\end{equation} 
where we have performed the integral with the formula in Appendix
\ref{identity}. Note that although the integrand in
Eq. (\ref{Uintimprov}) diverges as $v \to 0$, it does so very weakly
so that the
integral from $0$ to $\epsilon$ yields a contribution of
order $\epsilon \ln \epsilon$. Thus as before we expect the major
contribution to come from the region where $v$ is order $1$ and
$C_0(v{a\over r})$ is well approximated by $- 1/\ln(qr)$.

Differentiating this expression, and ignoring a term which is down in
magnitude by a factor $1/\ln(qr)$, (see the discussion following
 Eq.(\ref{approxTmunu})) we obtain the contribution
\begin{equation}
   {\bf F}_{\rm R}(r) =  - {\kappa Q^2 \over 16\epsilon_0 }
                              {\hat {\bf r} \over r^2 \ln(qr)}
   =  - {\kappa Q^2 \over 16\epsilon_0 }
                      {\hat {\bf r} \over r^2 \ln(2 e^{-{\cal C}}r/R)}.
\label{Fsuper}
\end{equation}
This is attractive, and, in the case that $R \sim a$, will be much
greater in magnitude than ${\bf F}_{\rm reg}$ over a very large 
range of $r$ values.  (We remark that for GUT scale strings, where
$\kappa$ is close to 1, it would be reasonable to replace $\kappa$ by
$1$ in (\ref{Uintimprov}) and (\ref{Fsuper}).) 

\begin{figure}
\centerline{\epsfig{figure=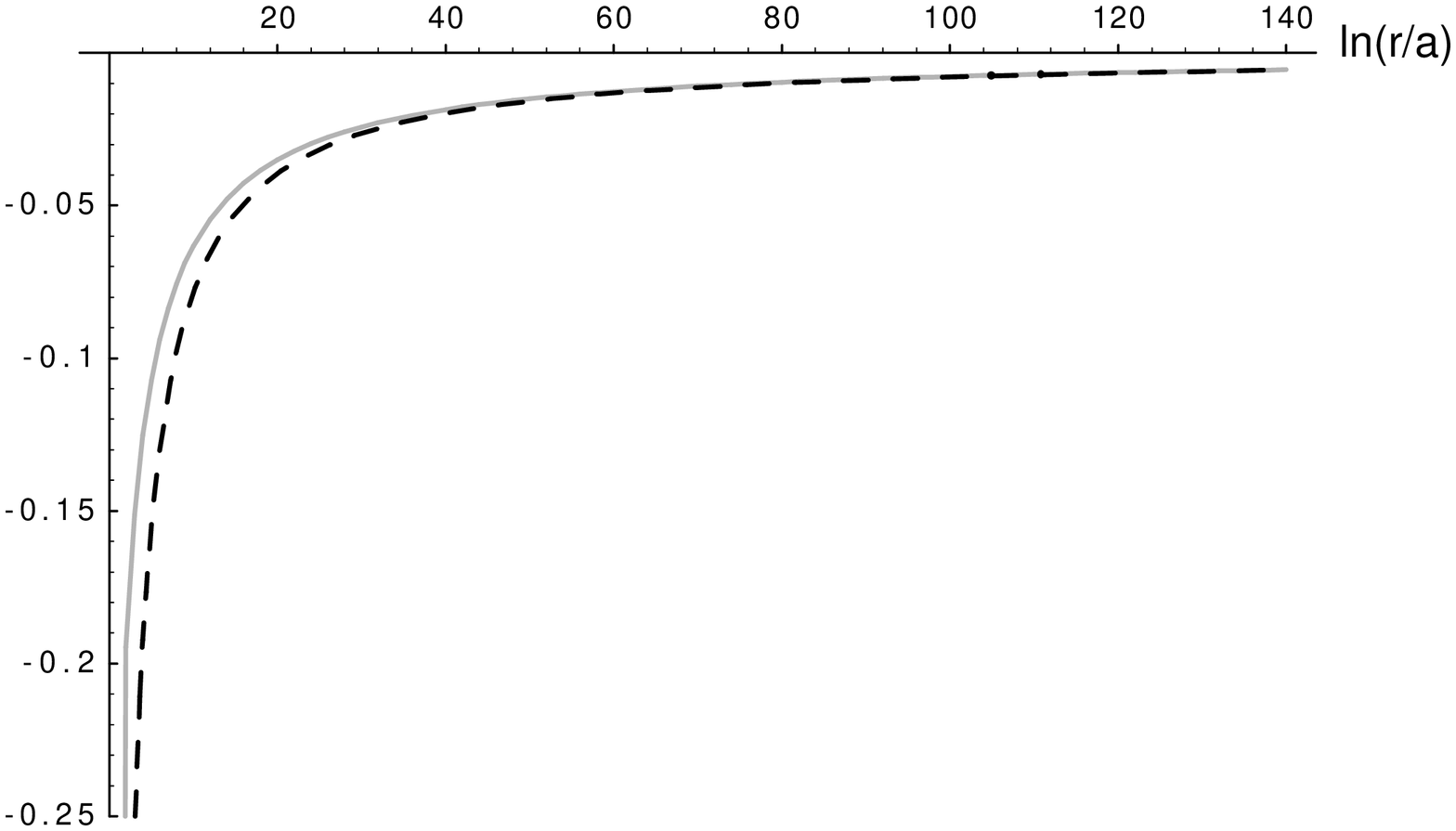,height=9truecm,
      bbllx=2.5truecm,bblly=1truecm
      ,bburx=16truecm,bbury=23truecm,angle=270} }
\caption{The solid curve is the (scaled) exact self-energy 
$W_{\rm super}(r) \times (4\pi
  \epsilon_0/Q^2) r$ and  the dashed curve is the (scaled) approximate
self-energy  $W_{\rm super}(r) \times (4\pi
  \epsilon_0/Q^2) r$  for perfect conductor boundary conditions on a
superconducting string with $\kappa = 100/99$.  The
gravitational contribution to the (scaled) self-energy is given by the
straight line $W_{\rm reg}  \times (4\pi
  \epsilon_0/Q^2) r = {\cal K}(100/99)/2 \approx  0{\cdot}0020 $.}
\label{selfenergyplot}
\end{figure}

As a simple model we may consider a potential $V$ given by
$V=\infty$ for $r<a$ and $V=0$ for $r>a$.  This ensures that
$\varphi$ vanishes at $r=a$ corresponding to a perfect conductor
boundary condition. In this case one immediately finds
\begin{equation}
      C_n(sa) = - {I_{\kappa |n|}(sa) \over K_{\kappa |n|}
(sa)} \label{perfectcn}
\end{equation}
and correspondingly $R=a$. In Fig. \ref{selfenergyplot} we plot
the exact correction to the renormalised self-energy, $W_{\rm super}$,
given by Eq. (\ref{Usuper}) and our corresponding approximation
$W_{\rm R}$ given by Eq. (\ref{Uintimprov}).

\section{CONCLUSION}
\label{conclusion}

There is an important distinction between the two calculations that we
have not yet mentioned.  The deviation of 
${\langle {\hat \varphi}^2 \rangle}$ from its ideal
value goes away if one imagines switching off gravity, that is, if the
deficit angle goes to zero, and the core becomes flat.  On the other
hand, for the self-force calculation, the deviation from the ideal
values does not go away if one ignores gravity.  In fact, the
gravitational contribution in this case is tiny in comparison with the
effect of the local photon mass term and hence we may ignore
gravitational effects and take the spatial metric to be flat in this case.

In conclusion, the calculations presented in this paper serve to
illustrate an
important general point of principle, namely:  the long-range
effects of cosmic strings (and more generally of `small objects'
\cite{KF}) can sometimes depend on the details of the structure of the
string core.

\acknowledgements
This work has been supported in part by Forbairt Grant No. SC/94/204,
NSF Grant No. PHY95-07740,  EPSRC Grant No. GR/K 29937,
 Forbairt Grant No. SC/94/204,
 and by a NATO Collaborative Research Grant.

\appendix
\section{THE KREIN RESOLVENT FORMULA}
\label{krein}

In this Appendix, we sketch the justification of our claim that the
expression (\ref{correction}) for the Green function 
$G_R^{(3)}({\bf x}, {\bf x'})$
-- when supplemented with a suitable ``principal part prescription''
-- is the exact Green function for the approximate Green function
equation (\ref{3Reqn})

\begin{equation}
-\triangle_R G_R^{(3)}({\bf x}, {\bf x'}) =
\delta^{(3)}({\bf x}, {\bf x'}).  
\end{equation}

We recall that $\triangle_R=\partial^2/\partial z^2 +
\triangle_R^{(2)}$ where $\triangle_R^{(2)}$ is the self-adjoint
extension of the two-dimensional Laplacian on the cone with deficit 
angle $2\pi(1-1/\kappa)$ corresponding to the boundary condition 
\begin{equation}
\varphi_R^{(n=0)}(r) \sim {\rm const}\cdot\ln(r/R) \qquad {\rm as}
\quad r \to 0,
\end{equation}
where $\varphi_R^{(n=0)}$ represents the $n=0$ sector component (i.e.
circular average) of an (elsewhere smooth) element of the domain of
$\triangle_R^{(2)}$.  (See \cite{KS} for a fuller discussion.)

Writing, ${\bf x}=(z,{\bf y})$, where ${\bf y}=(r,\phi)$ represents
a point on the 2-dimensional cone, formally, we clearly have
\begin{equation}
G_R^{(3)}({\bf x}, {\bf x'}) = {1 \over 2\pi}\int G_R^{(2)}({\bf y}, {\bf
y'}, k) e^{ik(z-z')}\, {\rm d}k
\end{equation}
where $G_R^{(2)}$ (which will, of course, be the Fourier transform of 
$G_R^{(3)}$ with respect to $z-z'$) satisfies
\begin{equation}
(-\triangle_R^{(2)} + k^2)G_R^{(2)}({\bf y},{\bf y'},k)=\delta^{(2)}
({\bf y},{\bf y'})
\end{equation}
together with the boundary conditions
\begin{equation}
G_R^{(2)}({\bf y},{\bf y'},k) \sim {\rm const} \cdot\ln(r/R) \qquad {\rm as}
\quad r \to 0,
  \label{krein1}
\end{equation}
for fixed ${\bf y'}$ and all $k$ (on which the `constant' may depend).
We now observe that the above conditions amount to the
statement that $G_R^{(2)}$ is the resolvent kernel of the self adjoint
extension $-\triangle_R^{(2)}$ of the two-dimensional cone Laplacian.
We may calculate this by {\it Krein's resolvent formula}.  (See for
example Appendix A in \cite{Alb}.)

This states (in the case of deficiency indices $\langle 1,1\rangle$)
that, given a symmetric operator $A$ (on a dense domain in some
Hilbert space) if $A_1$ and $A_2$ are a pair of its self-adjoint
extensions, then the difference in their resolvents is given by the
formula 
\begin{equation}
(A_1 - \lambda)^{-1} - (A_2 -
\lambda)^{-1}=f(\lambda)P_{\phi(\lambda)}         \label{krein2}
\end{equation}
where
\begin{itemize}

\item{}{(a)} $\lambda$ belongs to the resolvent set of each operator

\item{}{(b)} $\phi(\lambda)$ denotes a non-zero solution to 
\begin{equation}
A^*\varphi(\lambda)=\lambda\varphi(\lambda)           \label{krein3}
\end{equation}
with $A^*$ the adjoint of $A$, and $P_{\varphi(\lambda)}$ the projector
onto the subspace spanned by $\varphi(\lambda)$, and

\item{}{(c)} $f(\lambda)$ is an appropriate function to be fixed (see below).
\end{itemize}
If we identify $A$ with $-\triangle_R^{(2)}$ and $\lambda$
with $k^2$, then Eq.~(\ref{krein2}) is easily seen to be solved by
$\varphi(\lambda)=K_0(|k|r)$, so we conclude
that our resolvent kernel $G_R^{(2)}$ is related to the corresponding
kernel with regular boundary conditions by:
\begin{equation}
G_R^{(2)}({\bf y},{\bf y'},k)=G_{\rm reg}^{(2)}({\bf y},{\bf y'},k) 
+ f_R\left(s\right)K_0\left(sr\right)K_0\left(sr'\right)
\label{krein4}
\end{equation}
for some function $f_R\left(s\right)$, where as before $s=|k|$. 
Now
\begin{equation}
G_{\rm reg}^{(2)}({\bf y},{\bf y'},k) =
{\kappa \over 2\pi} \sum\limits_{n=-\infty}^\infty
    e^{in\Delta\phi}  I_{\kappa |n|}\left(sr_<\right)
               K_{\kappa |n|}\left(sr_>\right) ,
\end{equation}                       
so that $ G_{\rm reg}^{(2)}$ satisfies
\begin{equation}
  \label{regasymp}
  G_{\rm reg}^{(2)}({\bf y},{\bf y'},k)  \sim {\kappa \over 2\pi}
              K_0\left(sr'\right) \qquad {\rm as}
\quad r \to 0.
\end{equation}
Hence, from Eq.~(\ref{krein4})
 \begin{equation}
  \label{Rasymp}
  G_R^{(2)}({\bf y},{\bf y'},k)  \sim \left\{ {\kappa \over 2\pi}
             - f_R\left(s\right) \ln \left({e^{\cal C}s \over 2}r
             \right) \right\}
           K_0\left(sr'\right) \qquad {\rm as}
\quad r \to 0.
\end{equation}
To obtain agreement with Eq.~(\ref{krein1}) we must take
\begin{equation}
  {e^{\cal C} s  \over 2} \exp\left( - {\kappa \over 2 \pi}
     { 1 \over f_R\bigl(s\bigr)} \right) = {1 \over R} ,
\end{equation}
that is,
\begin{equation}
f_R(s)={\kappa\over 2\pi}\left[\ln\left(s/q\right)\right]^{-1}.
\end{equation}

Finally, multiplying both sides of (\ref{krein4}) by $(1/2\pi)e^{ik(z-z')}$
and integrating, we obtain (\ref{gR}).  We remark that, because of the
pole in $f_R\left(s\right)$ at $s=q$, 
the integral has to be interpreted as a
principle part integral.  It is not difficult to see that, when so
interpreted, the formula (\ref{gR}) does indeed yield a Green function
which satisfies (\ref{3Reqn}), i.e. which satisfies $-\triangle
G_R^{(3)}({\bf x}, {\bf x'})=\delta^{(3)}({\bf x}, {\bf x'})$ together
with the boundary condition
\begin{equation}
G_R^{(3)}({\bf x},{\bf x'},k) \sim {\rm const}\cdot \ln(r/R)   .
\end{equation}

\section{A USEFUL IDENTITY}
\label{identity}
  In deriving the approximate expressions for
 ${\langle {\hat \varphi}^2 \rangle}$ and 
${\langle {\hat T}_{\mu}{}^{\nu} \rangle}$
given in the text we have made frequent use of the identity
\begin{eqnarray}
\int\limits_0^\infty {\rm d}v \>&v^\lambda& K_\mu(v) K_\nu(v) =
   {2^{\lambda-2} \over \Gamma(1+\lambda)} \times \nonumber\\
 &&  \Gamma\left({1+\lambda+\mu+\nu \over 2}\right)
  \Gamma\left({1+\lambda-\mu+\nu \over 2}\right)
  \Gamma\left({1+\lambda+\mu-\nu \over 2}\right)
  \Gamma\left({1+\lambda-\mu-\nu \over 2}\right)\label{app}
\end{eqnarray}
valid for ${\Re}(\lambda) > |{\Re}(\mu)| + |{\Re}(\nu)|-1$.  
This equation may be 
readily derived from Eq.(6.576.4) of Gradsteyn \& Ryzhik\cite{GR}.

\begin{references}
\bibitem{footnote} 
In terms of the notion of self-adjoint extensions,
the Laplace-Beltrami operator $-\Delta$ for a constant-time
3-dimensional spatial section of the idealized string metric, when
defined on the domain of smooth functions compactly supported away
from line $r=0$ in the space of square-integrable functions on the
spatial section, has a one-parameter family of $z$-translationally
invariant self-adjoint extensions (corresponding to unrestricted
self-adjoint extensions for the 2-dimensional Laplace-Beltrami
operator on sections of constant $t$ and $z$) which are labelled by
the parameter $R$ taking its values in $[0,\infty)$, with the case
$R=0$ corresponding to regular boundary conditions, and non-zero $R$
to the behaviour (\ref{kayasymp}).
\bibitem{AOcorr} 
Note that only the leading asymptotic form for $C_0(x)$ for the 
`flower-pot' was given in Ref.\cite{AO}.  The 
next term which is important here was given in  
B.~Allen, J.G.~McLaughlin and A.C.~Ottewill, Phys.Rev.D {\bf 45}, 4486 
(1992).
\end{references}
\end{document}